\documentclass[modern]{aastex631}
\usepackage{amssymb, amsmath}
\usepackage{color}
\usepackage{graphicx}
\usepackage{morefloats}
\usepackage{hyperref}
\NewPageAfterKeywords
\begin{document}

\title{The Discovery of Three Galactic Wolf-Rayet Stars}

\author[0009-0008-3389-9848]{Laurella C. Marin}
\affiliation{Lowell Observatory, 1400 W Mars Hill Road, Flagstaff, AZ 86001, USA}
\email{laurella.c.marin.25@dartmouth.edu}

\author[0000-0001-6563-7828]{Philip Massey}
\affiliation{Lowell Observatory, 1400 W Mars Hill Road, Flagstaff, AZ 86001, USA}
\affiliation{Dept. of Astronomy \& Planetary Science, Northern Arizona University, PO Box 6010, Flagstaff, AZ 86011, USA}
\email{massey@lowell.edu}

\author{Brian A. Skiff}
\affiliation{Lowell Observatory, 1400 W Mars Hill Road, Flagstaff, AZ 86001, USA}

\author[0000-0001-5750-4953]{Kennedy A. Farrell}
\affiliation{Lowell Observatory, 1400 W Mars Hill Road, Flagstaff, AZ 86001, USA}
\affiliation{Dept. of Astronomy \& Planetary Science, Northern Arizona University, PO Box 6010, Flagstaff, AZ 86011, USA}

\begin{abstract}
Wolf-Rayet stars (WRs) are evolved massive stars in the brief stage before they undergo core collapse. Not only are they rare, but they also can be particularly difficult to find due to the high extinction in the Galactic plane. This paper discusses the discovery of three new Galactic WRs previously classified as H$\alpha$ emission stars, but thanks to Gaia spectra, we were able to identify the broad, strong emission lines that characterize WRs. Using the Lowell Discovery Telescope and the DeVeny spectrograph, we obtained spectra for each star. Two are WC9s, and the third is a WN6 + O6.5~V binary. The latter is a known eclipsing system with a 4.4 day period from ASAS-SN data. We calculate absolute visual magnitudes for all three stars to be between -7 and -6, which is consistent with our expectations of these subtypes. These discoveries highlight the incompleteness of the WR census in our local volume of the Milky Way and suggest the potential for future Galactic WR discoveries from Gaia low-dispersion spectra.  Furthermore, radial velocity studies of the newly found binary will provide direct mass estimates and orbital parameters, adding to our knowledge of the role that binarity plays in massive star evolution.

\end{abstract}

\section{Introduction}

Wolf-Rayet stars (WRs) are the evolved descendants of massive O-type stars and are in the last stage of evolution before they undergo core collapse. (For a recent review, see \citealt{2019Galax...7...74N}.) As such, they are rare. Recent surveys have identified 158 WRs in the Large Magellanic Cloud, 12 in the Small Magellanic Cloud, and 206 WRs in M33 \citep{2011ApJ...733..123N, 2018ApJ...863..181N}, numbers that are likely complete. There are currently 173 WRs known in M31 \citep{2012ApJ...759...11N,2023AJ....166...68N}, although the true number is estimated to exceed 200, as dust obscuring the disk of M31 has made it difficult to achieve a complete census \citep{2023AJ....166...68N}. 

The extinction problem is worse for surveys in our own Galaxy. The first modern catalog of WRs in the Milky Way by \citet{1981SSRv...28..227V} listed 159 WRs. This catalog was revised in 2001, and the number of WRs increased to 227 \citep{2001NewAR..45..135V}. Thanks to new near-infrared surveys, the online list of Galactic WRs\footnote{https://pacrowther.staff.shef.ac.uk/WRcat/}, maintained by Paul Crowther at the University of Sheffield, lists 676 as of the latest update in May 2024. (For a relatively recent review of the Wolf-Rayet content of the Milky Way, see \citealt{2015wrs..conf...21C}.) 

Complete censuses of massive star populations provide the means to conduct important tests of stellar evolutionary theory.  Examples of such tests that have been conducted using our knowledge of stellar populations in the Magellanic Clouds, M33, and M31 include the luminosity functions of yellow supergiants \citep{2009ApJ...703..441D,2010ApJ...719.1784N,2012ApJ...750...97D}, the luminosity function of red supergiants (RSGs) \citep{2020ApJ...889...44N,2023ApJ...942...69M}, the binary fraction of RSGs and its implication for merger rates \citep{2020ApJ...900..118N,2021ApJ...908...87N} and the relative number of WRs and RSGs \citep{2021ApJ...922..177M}. Such tests are difficult to carry out in the Galaxy. While obtaining a galaxy-wide census of Milky Way WRs will likely never be possible due to the insurmountable observational challenges, identifying WRs in a local volume-limited sample provides the means for similar tests. \citet{1983ApJ...274..302C} presents an example of such using O-type and WR stars in a sample within 2.5~kpc of the Sun, believed at the time to be complete.

\begin{figure}[h!]
\epsscale{0.53}
\plotone{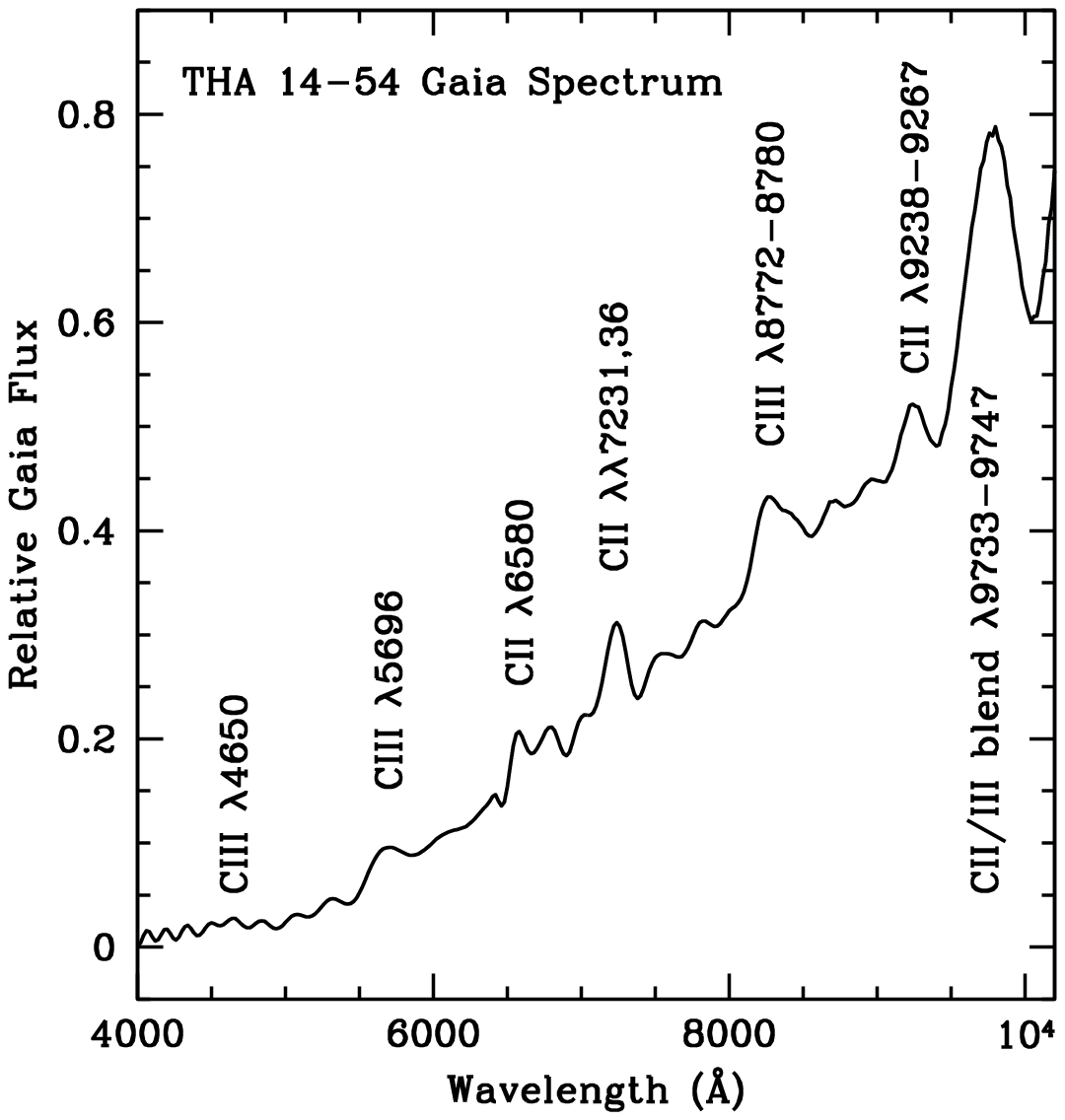} 
\plotone{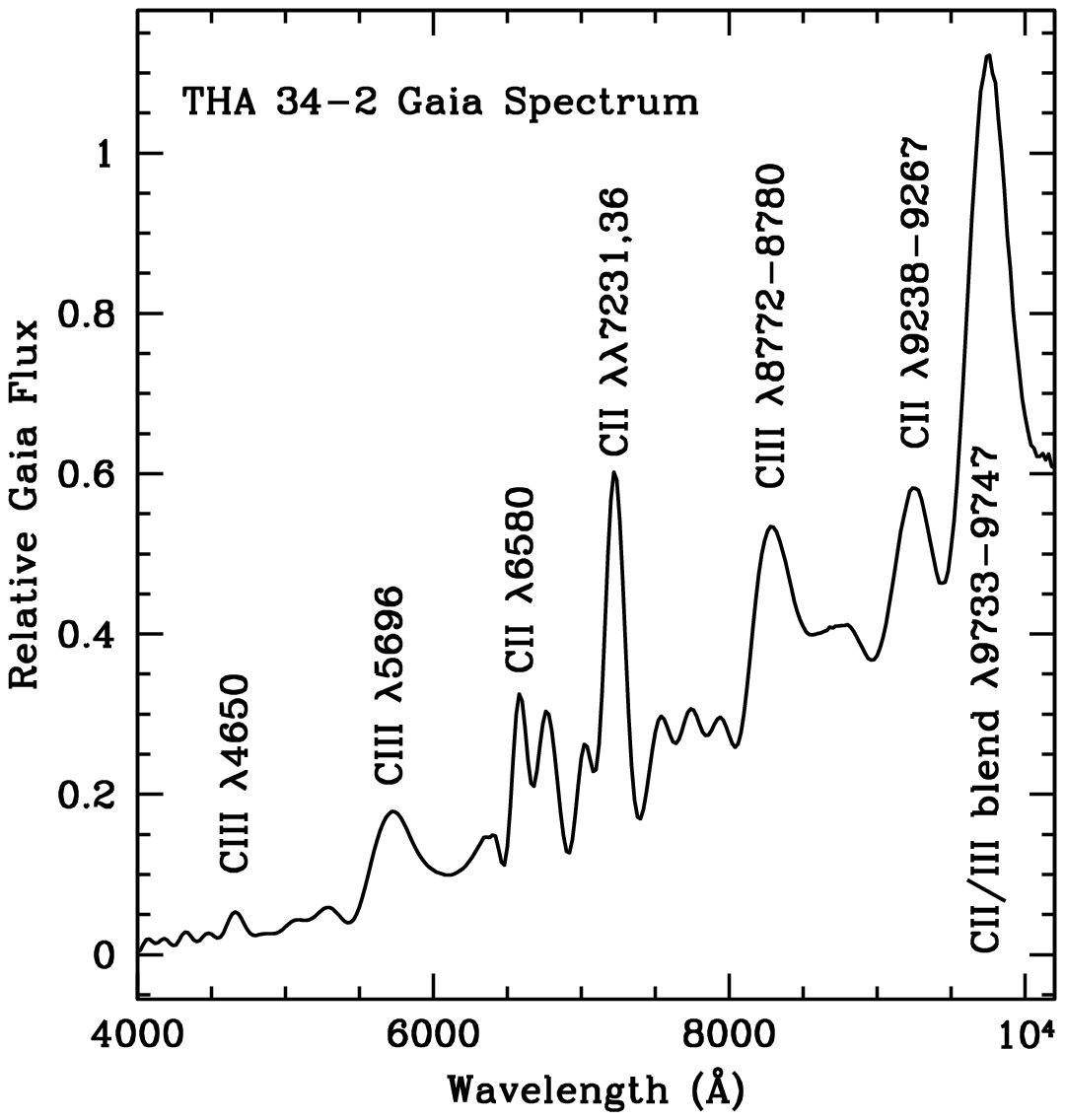}
\plotone{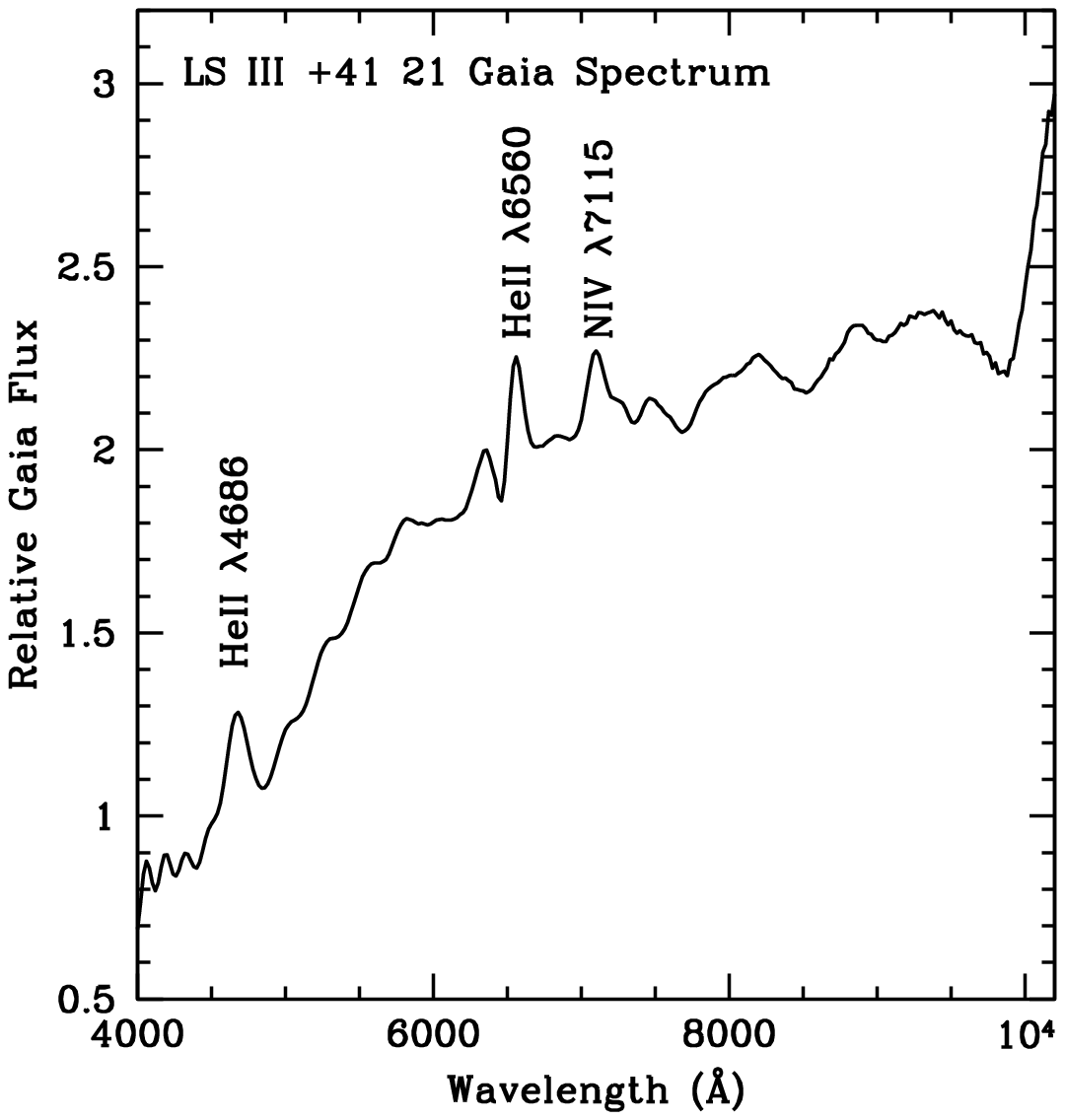}

    \caption{Gaia spectra of our three newly found WRs.  These sources were initially classified as H$\alpha$ sources
    \citep{1962CoBos..14....0T,1966CoBos..34....1T,1964LS....C03....0H} from objective prisms surveys.  Our inspection of the Gaia spectra showed the broad, strong lines indicative of WR stars.}
    \label{fig:gaia}
\end{figure}

In this paper, we describe follow-up spectroscopy of three relatively bright, previously unknown Galactic WRs: THA 14-54, THA 34-2, and LS III +44 21. 
The three stars were first noted as possible H$\alpha$ emission-line sources by \citet{1962CoBos..14....0T}, \citet{1966CoBos..34....1T}, and \citet{1964LS....C03....0H}, respectively.   In order to follow up such older, purported H$\alpha$ sources,  one of us (B.A.S.) made use of Gaia spectra that had recently been made available as part of its Data Release 3 (DR3)  \citep{2016A&A...595A...1G,2023A&A...674A...1G}.  These spectra are of low-resolution (resolving power $R\sim 30-100$) and provide wavelength coverage from  3300~\AA\ to 10500~\AA\ of 219 million sources (\citealt{2023A&A...671A..52W} and references therein). Inspection showed that many of these alleged H$\alpha$ sources in fact had no emission, but three of them showed the broad, strong emission lines characteristic of WRs.  Two were clearly of late WC type and one of WN type, as shown by comparison with low-dispersion far-red spectra shown in \citet{1990ApJ...354..359C}.  These three Gaia spectra are shown in Figure~\ref{fig:gaia}.  We report here the results of higher dispersion spectroscopy and analysis of these three newly found WRs.

In Section~\ref{Sec-obs}, we describe our observations and reductions.  In Section~\ref{Sec-class} we classify these newly discovered WRs. In Section~\Ref{Sec-props}, we give extinction estimates and absolute magnitudes utilizing Gaia distance determinations. In Section ~\Ref{Sec-discussion} we provide the main takeaways from these discoveries and give recommendations for future research. 

\clearpage

\section{Observations and Reductions}
\label{Sec-obs}

We obtained spectra of these three stars using the 4.3 m Lowell Discovery Telescope (LDT) located near Happy Jack, AZ. The DeVeny optical spectrograph was used for all observations. The detector is a 2048 $\times$ 512 e2v CCD42-10 deep-depletion device with 13.5 $\mu$m pixels. It has a full-well capacity of about 100,000 electrons and is operated at a gain of 1.5 electrons ADU$^{-1}$. A Stirling closed-cycle cooler chills the chip to $-110^\circ$C, and a slit-viewing camera is used to position stars on the slit. The spatial scale on the LDT is 0\farcs34 pixel$^{-1}$.

Observing time had been assigned on the LDT on (UT) 21 and 22 June 2024 in connection with a different project, and this provided an opportunity to observe all three of these stars in the blue. Conditions on both nights were very poor, with seeing $>$2\arcsec. On 21 June we observed THA 34-2 and THA 14-54 as well as the spectrophotometric standard Wolf 1346. Exposure times for the WR star were 600~s plus 3$\times$1200~s for THA 34-2, and 2$\times$1200 s for THA 14-54. We planned for more exposure on the latter but closed early due to humidity. Conditions on 22 June were even worse, with thick clouds, but we managed to obtain good signal on our third star, LS III +44 31, in a few minutes of workable conditions. We were intrigued after a looking at the spectra, so we decided we wanted to extend coverage to the red. We were kindly granted an hour out of engineering time assigned on UT 5 July to do so.  

A 600 line mm$^{-1}$ grating (DV6) blazed at 4900~\AA\ was used for the blue observations, providing a wavelength coverage of 3720--6000 \AA. The dispersion was 1.12~\AA\ pixel$^{-1}$, and with the slit opened to 2\farcs5, we achieved a spectral resolution of 5.4~\AA. 
%The grating tilt was $27.04^\circ$ for the first two nights. 
A 400 line mm$^{-1}$ grating (DV4) blazed at 8500~\AA\ was used for the red observations on the third night, covering a wavelength range of 5839--9230 \AA. The dispersion was 1.67~\AA\ pixel$^{-1}$, and with the same slit width, the spectral resolution was 8.3~\AA. 
%The grating tilt was $27.37^\circ$ on the third night. 
We used an OG570 filter to block out second-order blue light for the red observations. We obtained all stellar observations with the slit oriented to the parallactic angle.

To wavelength calibrate the spectra, we took comparison exposures of Hg-Cd-Ar lamps each night we observed in the blue and of a Ne comparison lamp for the red observations. These were obtained at zenith. Although the spectrograph does experience some flexure (a few pixels), the procedure for taking comparison exposures is too time-consuming to put on comparison exposures at each sky position. In any event, we were interested in the wavelength solution primarily for line identifications.   

We reduced the data using standard {\sc iraf} routines. We measured the overscan and then subtracted from each image. The images were trimmed to 2041$\times$387 pixels for the first two nights and 2042$\times$300 pixels for the third night. Biases were taken, but they were not used on any night because no structure was identified in the images. We took dome flats for each grating setting. These were then averaged, normalized, and divided into the science and comparison arc frames. 

A $\pm$15 pixel extraction aperture was defined around the peak of the star for the blue spectra, and a $\pm$18 pixel aperture was used for the red observations. The position of the spectrum on the CCD was traced, and an optimal extraction algorithm using pixel rejection was applied along that trace. The comparison spectra were extracted using the same traces on a star-by-star basis and fit with a low-order cubic spline that resulted in residuals of about 0.04~\AA.  

The standard star observation was then used to flux calibrate the blue spectra for the purposes of measuring the reddening, as described in Section~\ref{Sec-props}. For this, we adopted the ``standard" Kitt Peak extinction terms as a function of wavelength.  Although the data were obtained under poor conditions, especially for LS III +44 21, our experience is that clouds are fairly grey, and we expect that the {\it relative} flux calibration is good to 5\%. In addition, both the blue and the red spectra were normalized and combined to obtain complete coverage for each star from 3720~\AA\ to 9230~\AA\ for line identification and illustration in this paper.

\section{Spectral Classifications}
\label{Sec-class}

In this section, we assign spectral subtypes to our three newly found WRs. The spectra of WN-type WRs are dominated by helium and nitrogen lines, which are the products of the hydrogen-burning CNO cycle. Their classification depends primarily upon the relative strengths of N\,{\sc iii}, N\,{\sc iv}, and N\,{\sc v} emission lines. At later types (WN7-11), the relative strengths of the N\,{\sc iii} lines are compared to He\,{\sc ii} $\lambda 4686$, and we look for the presence of N\,{\sc ii}. The WC-type stars are more chemically evolved than WNs; their spectra are dominated by the products of helium burning: carbon and oxygen. Their subtypes are determined by the relative strengths of C\,{\sc ii}, C\,{\sc iii}, C\,{\sc iv}, and O\,{\sc v} emission lines. The relationship between the subtypes is an area of active research \citep{2007ARA&A..45..177C}. We know that the relative number of WCs and WNs depends upon the metallicity of the regions in which the stars formed, with a larger fraction of WCs found in regions of higher metallicity (see, e.g., \citealt{1998ApJ...505..793M, 2019Galax...7...74N}, and references therein). Similarly, later-type (lower excitation) WC stars (WC7-9) are found in regions of higher metallicity, as first noted by \citet{1968MNRAS.141..317S} and subsequently confirmed in other galaxies (e.g., \citealt{2011ApJ...733..123N}).

We classified each newly discovered WR using the qualitative criteria that was originally described by \citet{1968MNRAS.138..109S}, subsequently updated by others, and summarized in Table 2 of the \citet{2001NewAR..45..135V} VII Catalogue of Galactic WR Stars.
\begin{figure}[h!]
\epsscale{1}
\plotone{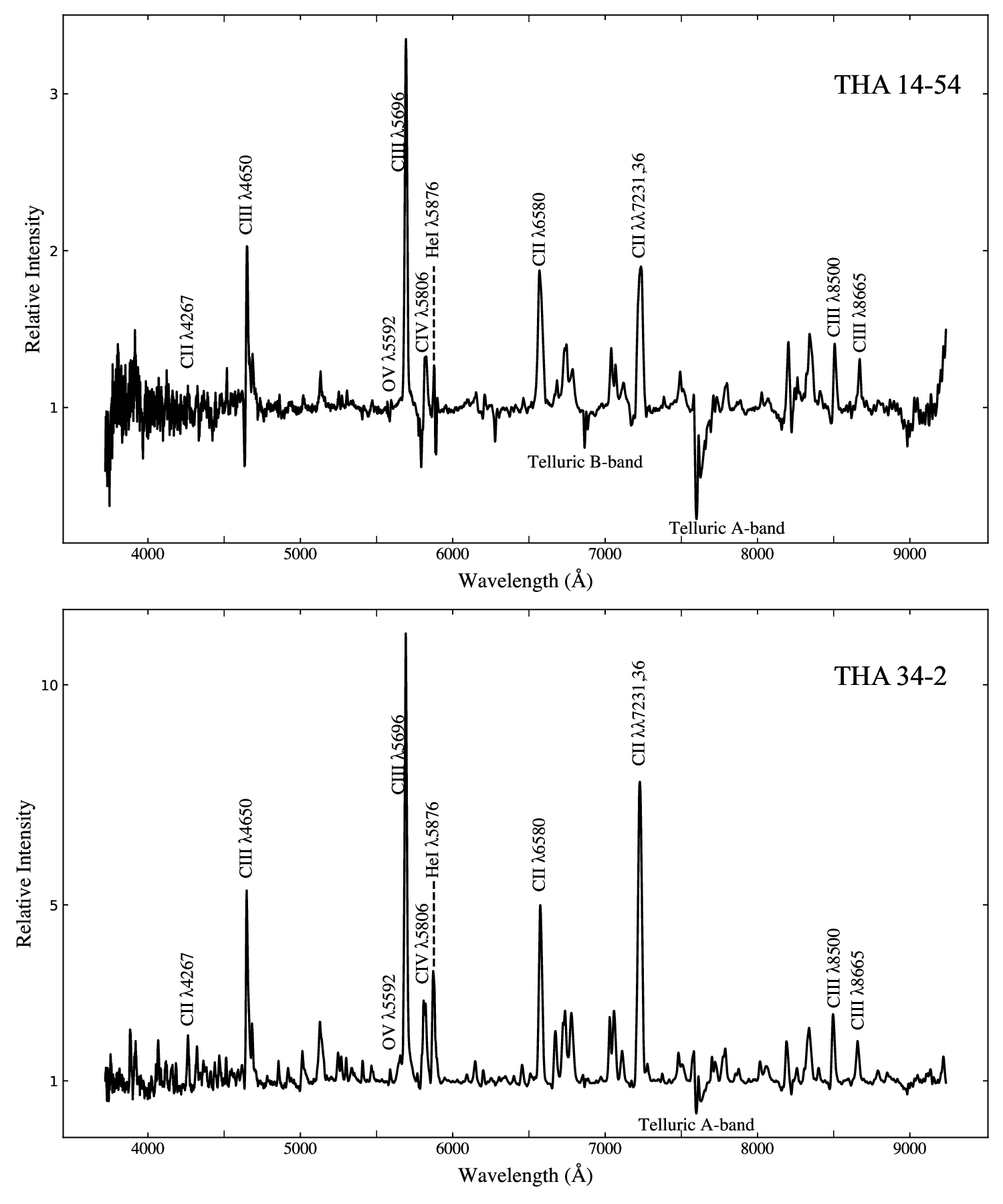} 
    \caption{The spectra of the two WC9 stars. The classification lines along with other strong lines are labeled. It may be instructive to compare these spectra to the lower resolution Gaia spectra shown in Figure~\ref{fig:gaia}.}
    \label{fig:WC9s}
\end{figure}

\subsection{Two New WC9s}
THA 34-2 and THA 14-54 are each classified as WC9. In order to classify WCs, the relative strengths of the lines C\,{\sc iv} $\lambda \text{5806}$, C\,{\sc iii} $\lambda \text{5696}$, and O\,{\sc v} $\lambda \text{5592}$ are used. In both of these stars, we found that C\,{\sc iii} was stronger than C\,{\sc iv}, and O\,{\sc v} is only weakly present. Furthermore, C\,{\sc ii}  $\lambda \text{4267}$ and C\,{\sc ii} $\lambda \lambda$ 7231,36 are strong in both. The presence of C\,{\sc ii} eliminated the possibility of these stars being WC8s and thus confirmed that both are WC9s.  Figure~\ref{fig:WC9s} shows the spectra of the two WC9s with the principal lines identified. 

Note that although both of these stars were once considered to be H$\alpha$ sources, neither has the H$\alpha$ line nor the corresponding Pickering He\,{\sc ii} line at 6560~\AA.  Instead, the strong C\,{\sc ii} $\lambda$6580 doublet was doubtless mistaken for H$\alpha$.

\subsection{A Newly Found WN+O Binary}

\begin{figure}[h!]
    \epsscale{1}
    \plotone{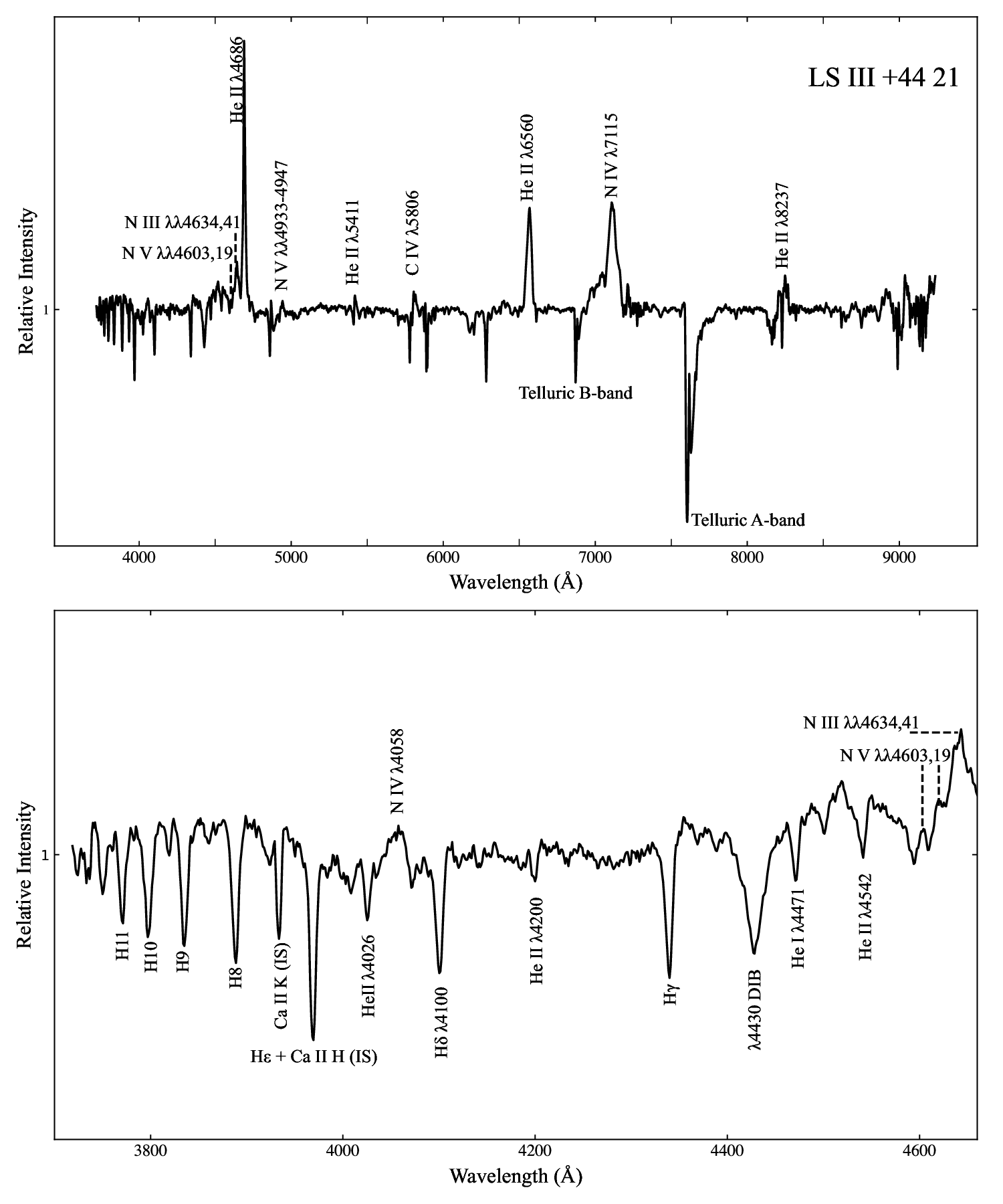}]
    \caption{The spectra of the WR binary LS III +44 21. The bottom panel is zoomed in to the bluest part of the spectrum to show the absorption lines. The strongest interstellar (IS) lines have been identified, including the 4430\AA\ diffuse interstellar band (DIB). Compare to the lower resolution Gaia spectra shown in Figure~\ref{fig:gaia}.}
    \label{fig:LS}
\end{figure}

We classify LS III +44 21 as WN6 with an O6.5~V companion.  
The WN subtypes depend upon the relative strengths of  N\,{\sc iii} $\lambda \lambda$ 4634, 42, 
N\,{\sc iv} $\lambda$4058, N\,{\sc v} $\lambda \lambda$ 4603,19, and 
He\,{\sc ii} $\lambda$4686.  For this star, N\,{\sc iii} is roughly the same strength as 
N\,{\sc iv} but much weaker than He\,{\sc ii}, and  N\,{\sc v} is only weakly present 
(Figure~\ref{fig:LS}).

The presence of He\,{\sc ii} $\lambda \text{4542}$ and He\,{\sc ii} $\lambda \text{4200}$ absorption indicates that the companion star must be an O-type star. Using  the classification criteria displayed in \citet{1990PASP..102..379W}, we classified the companion as an O6.5; i.e., He\,{\sc i} $\lambda$4471 is just slightly weaker than He\,{\sc ii} $\lambda$4542. (See Figure~\ref{fig:LS}.) To determine whether the companion is a supergiant or a dwarf, we compared the strength of the Si\,{\sc iv} $\lambda \text{4089,4116}$ lines to the H$\delta$ absorption, and we conclude it is likely a dwarf. This method as used as an alternative to the He\,{\sc ii} $\lambda$4686 line strength (where emission indicates a supergiant, while absorption indicates a dwarf) since that line is dominated by emission from the WR star. Thus we conclude the companion is an O6.5~V. 

Note that for LS III +44 21 the identification of the star as an H$\alpha$ source was {\it nearly} correct. The line is actually the corresponding Pickering He\,{\sc ii} line at 6560\AA.

\section{Stellar Properties}
\label{Sec-props}

\subsection{Reddenings and Absolute Magnitudes}

One of the most fundamental properties of a star is its absolute visual magnitude. For WRs, this quantity is somewhat less fundamental because the V-band is dominated by strong emission lines originating from the stellar wind; their strengths are, dependent upon mass-loss rates which are tied to temperature and luminosity in a complicated manner.  The presence of emission lines throughout the B- and V-bands further complicates attempts to measure reddenings, as the intrinsic $(B-V)_0$ values depend on line strengths and subclass. To address this issue, \citet{1966ApJ...145..724W} introduced a set of narrow-band filters (denoted by lower-case letters, i.e., $b$ and $v$, each roughly 100~\AA\ wide) designed to avoid emission lines as much as possible. \citet{1984ApJ...281..789M} noted that for many subtypes—particularly WCs—even these narrow-band interference filters contained emission.  The advent of digital detectors (including eventually CCDs) in the 1980s moved spectrophotometry away from specialized instruments onto conventional spectrographs. \citet{1984ApJ...281..789M} took advantage of this new ability to measure monochromatic magnitudes at the $b$ and $v$ central wavelengths in order to determine reddenings and absolute magnitudes of WRs.  (See also \citealt{1988AJ.....96.1076T}.)

However, here we are partially stymied.  Our own fluxed data are good enough for measuring colors, but given the observing conditions, they cannot be relied upon for absolute values.  Broad-band V magnitudes however are available from the literature.  So, in this section we will derive $M_V$ values rather than $M_v$, but will rely upon $b$ and $v$ for determining the correction for interstellar extinction.

We measured the flux of each star at 4270~\AA\ (``$b$") and 5160~\AA\ (``$v$") following \citet{1984ApJ...281..789M}. The \citet{1966ApJ...145..724W} system is tied to the spectrophotometric AB system, i.e., stars with constant $f_\nu$ have zero colors. To determine the reddening we assumed that $(b-v)_0$ = -0.30 based upon Table 5  in \citet{1984ApJ...281..789M}. We then adopted $E(B-V)=1.20\times E(b-v)$ from Table VIII in \citet{1968MNRAS.140..409S}.  We then adopted the standard ratio of total to selective extinction ratio $R_V=3.1$ to compute $A_V$.  Thus $A_V = 3.72 E(b-v)$.  We obtained $V$ values from the literature as given in Table~\ref{tab:NewWRs}, and computed $V_0=V-A_V$.  Note that for the two WC9s the $A_V$ values are 7-7.5~mags;
as shown in Figure~\ref{fig:gaia} these stars are indeed heavily reddened. 

For the distances, we relied upon the Gaia parallax determinations given by \citet{2021AJ....161..147B} and their 16th and 84th percentile estimates were adopted as their uncertainties. The distance moduli and $V_0$ values were then used to calculate the final absolute visual magnitudes; the values are given in Table~\ref{tab:NewWRs}.
\begin{deluxetable}{l c c c c c l l l}
\tabletypesize{\scriptsize}
\tablecaption{\label{tab:NewWRs}Three Newly Confirmed Galactic Wolf-Rayet Stars}
\tablewidth{0pt}
\tablehead{
\colhead{Star}
&\colhead{$\alpha_{\rm 2000}$}
&\colhead{$\delta_{\rm 2000}$}
&\colhead{Sp.\ Type}
&\colhead{$V$}
&\colhead{$(b-v)$\tablenotemark{a}}
&\colhead{$A_V$}
&\colhead{Dist.\ (kpc)\tablenotemark{b}}
&\colhead{$M_V$} 
}
\startdata
 THA 34-2 & 18:04:52.43 & $-$20:37:48.6 & WC9 & 14.01\tablenotemark{c} & 1.59 & $7.0\pm0.2$ & 3.99$^{+0.6}_{-0.5}$ & $-6.0\pm0.5$ \\
 THA 14-54&18:40:14.97 & $-$05:03:20.0 & WC9 & 13.86\tablenotemark{d} & 1.73 & $7.5\pm0.2$ & 4.33$^{+1.33}_{-0.70}$ & $-6.8\pm0.8$ \\
 LS III +44 21 &  20:43:36.48 &+44:55:05.3 & WN6+O6.5~V & 10.89\tablenotemark{d} & 0.90 & $4.5\pm0.2$ & 3.36$^{+0.17}_{-0.15}$ & $-6.3\pm0.4$\\
 \enddata
 \tablenotetext{a}{Monochromatic flux AB magnitudes at 4270~\AA\ and 5160~\AA\ measured from our spectra; see e.g.,\citet{1984ApJ...281..789M}.}
 \tablenotetext{b}{Gaia distance from \citet{2021AJ....161..147B}.}
 \tablenotetext{c}{\citet{2013AJ....145...44Z}.}
 \tablenotetext{d}{\citet{2004AAS...205.4815Z}.}
  \end{deluxetable}
  
These absolute magnitudes are all consistent with those of Population I WRs, as expected. In their complete catalog of LMC WRs, \citet{2018ApJ...863..181N} include $V$ magnitudes. Adopting $A_V=0.4$ as typical of early-type stars in the LMC, and a distance of 50~kpc, leads to an apparent distance modulus of 18.9. We find that the non-binary WCs in their catalog typically range from $M_V=-5.8$ to $-4.3$, with a few brighter and fainter.  Of course, all of these WCs are of the WC4 subtype, and we know from Galactic studies that WC9s are typically 1-2 magnitudes brighter than early-type WCs (see, e.g., Figure 4 of \citealt{2001NewAR..45..135V}).  Thus absolute magnitudes of -6.0 (THA 34-2) and -6.8 (THA 14-54) are as expected.

From the \citet{2018ApJ...863..181N} table we find that three WN6 stars in the LMC have $M_V=-6.7$ to $-4.8$.  A typical O6.5~V has $M_V\sim-5.0$ \citep{1983ApJ...274..302C}.  So, finding $M_V=-6.3$ for LS III +44 21 is also consistent with expectations.

 \subsection{Are Our WC9s Dust Producers?}

WC9 stars have long been recognized as sources of dust; see, e.g., \citet{2015MNRAS.449.1834W} and references therein.
Although the vast majority of dust in the solar neighborhood comes from asymptotic giant branch (AGB) stars \citep{Dwek},  \citet{Smoke} notes that in the early universe, or in starbursts, both red supergiants and WC9 stars may be major contributors.
Such dust is evidenced by thermal excess at longer wavelengths ($K$-band and further).

There are WC9 stars that produce dust spectacularly  during periastron passage in long period eccentric orbits with a massive companion.  WR140 (HD 193793) is the classic example of such a system \citep{1990MNRAS.243..662W}. These are known as ``colliding wind" dust producers.  Other WC9s are {\it persistent} dust producers.\footnote{The notation ``WC9d" was introduced by \citet{2001NewAR..45..135V} in order to designate the persistent dust producing WC9s.  Although this nomenclature may be convenient, we find it unfortunate as it relies upon ``outside" information (i.e., IR excess)  as part of a spectral type designation, violating the basic morphological basis for spectral classification.}    In some cases, such as WR104, the mechanism is the same, as the star is a close binary system, but the orbit is circular.  Not all persistent dust producers are known to be binaries, and it is unclear if binarity is needed in order to persistently produce dust (see \citealt{2000MNRAS.314...23W} and \citealt{2003Ap&SS.285..677C}). Other WC9s,  such as WR81, WR88, and WR92 (HD 157451), show no signs of dust emission \citep{2000MNRAS.314...23W,2015MNRAS.449.1834W}.

It is of interest, then, to see if our newly found WC9 stars are persistent dust producers.\footnote{We thank the anonymous referee for posing this questiion.}  To answer this,  we de-reddened the stars' $V-K$ colors, assuming
that $E(V-K)\sim3.0E(B-V)$ \citep{Sch}, and adopting the crude approximation that 
$(B-V)_0=-0.3$.  (Since we are comparing relative $(V-K)_0$ colors the exact value is irrelevant.)  We find that most of the persistent dust producing WC9s listed in Table 1 of \citet{2000MNRAS.314...23W} have $(V-K)_0$ values between 2.5-6.5. (Specifically, we used WR53, WR95, WR96, WR104, and WR106.)  By contrast, the non-dust producers  WR81, WR88, and WR92 have $(V-K)_0$ colors of -0.6.  The division is not totally clean; for instance, the persistent dust producer WR103 has a $(V-K)_0$ color of only +0.7.  But generally we would consider a $(V-K)_0 > 2$ mags to be clear indicator of a dust producer.  By contrast, the $(V-K)_0$ of THA 34-2 is 1.3 and THA14-54 is -0.3.  At best, this is
suggestive that of some dust production by THA 34-2.

We have previously found that the extinction of these stars are large.  Could this be an indicator of dust causing circumstellar reddening?  The answer is no.  In general, the typical extinction in the plane of the Milky Way is 1.8 mag kpc$^{-1}$ \citep{2003dge..conf.....W}.  With distances of 4.0 and 4.3~kpc, extinctions of $A_V$ of 7.0 and 7.5~mag (Table~\ref{tab:NewWRs}) are just what we expect for THA 34-2 and THA 14-54.  There is no evidence of substantial circumstellar extinction due to dust.

%\clearpage

\section{Discussion}
\label{Sec-discussion}

WRs are rare, and the WC9s are particularly rare. Throughout the Local Group, they seem to be found only in regions where the metallicity is above solar. To find two WC9s less that 5 kpc away demonstrates that the local volume of the Milky Way has not been thoroughly explored. There are opportunities for many more discoveries of WRs and a long-term opportunity to establish a volume-limited census around the Sun. 

\begin{figure}[h!]
    \centering
    \includegraphics[width=\textwidth]{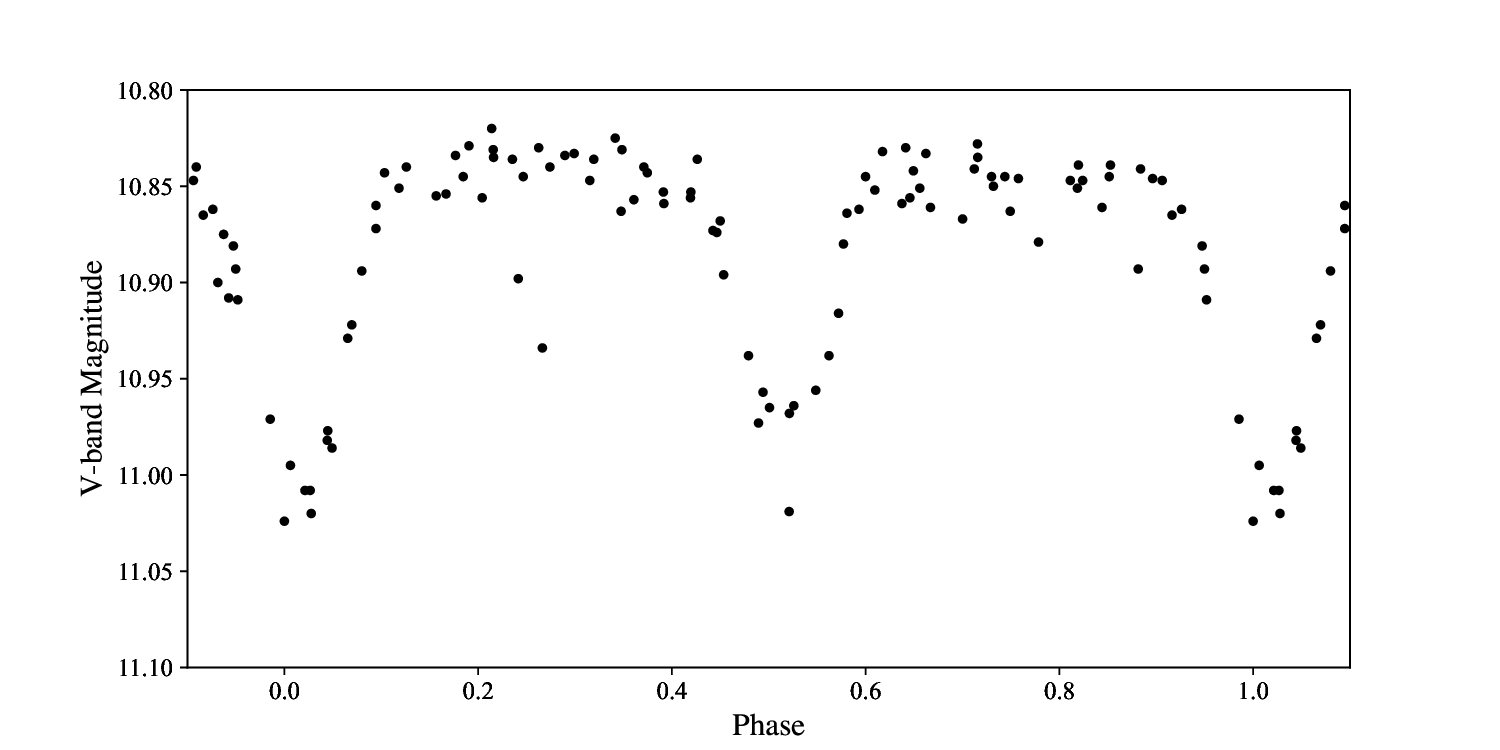} % Change the filename to your EPS file
    \caption{The V-band light curve of LS III +44 21 based on ASAS-SN data. We have phased the data using a period of 4.4313725 day and a time of primary eclipse of Julian date 2457270.94 based on the values given in \citet{Jayasinghe_2021}. We note that their webpage for this star indicates that some of these data may be saturated.}
    \label{fig:lightcurve}
\end{figure}

The discovery of an 11th magnitude WR binary just 3.4 kpc away further highlights the incompleteness of our knowledge of the massive star content of our neighborhood. This binary system has a period of 4.43 days according to \citep{Jayasinghe_2021}. The light curve shown in Figure~\ref{fig:lightcurve} shows a distinguishable primary and secondary eclipse, and was constructed from the online ASAS-SN dataset \citep{2014ApJ...788...48S,2017PASP..129j4502K}\footnote{https://asas-sn.osu.edu/variables/b3d640ee-93b6-5d6f-891f-7f7a07578853}. Radial velocities studies could determine masses for the two components; the orbital parameters would provide further insights regarding the role of binarity on the formation of WRs. 

The availability of low-dispersion spectra of 219 million stars from Gaia DR3 \citep{2023A&A...674A...1G,2023A&A...671A..52W} provides an invaluable resource for significantly expanding our knowledge of the massive star content of our neighborhood. The spectral signatures of WR stars are easily recognized from such data (see the example we show in Figure~\ref{fig:gaia}), and it is only a matter of time before the power of machine-learning and AI are brought to bear on the matter. Meanwhile, reviewing older papers of supposed H$\alpha$ emission sources could be a promising place to find other misclassified stars, particularly WC9s and WNs.

\section{Acknowledgements}
Lowell Observatory sits at the base of mountains sacred to tribes throughout the region. We honor their past, present, and future generations, who have lived here for millennia and will forever call this place home. 

We acknowledge useful comments by an anonymous referee which improved this paper. L.C.M.'s participation was supported through the Research Experiences for Undergraduates program funded through Northern Arizona University thanks to the National Science Foundation (NSF) grants 1950901 and 2349774.
P.M. and K.A.F. gratefully acknowledges support from the NSF through AST-2307594. 

These results made use of the Lowell Discovery Telescope (LDT) at Lowell Observatory.  Lowell is a private, non-profit institution dedicated to astrophysical research and public appreciation of astronomy and operates the LDT in partnership with Boston University, the University of Maryland, the University of Toledo, Northern Arizona University and Yale University.  The upgrade of the DeVeny optical spectrograph has been funded by a generous grant from John and Ginger Giovale and by a grant from the Mt. Cuba Astronomical Foundation.

We gratefully acknowledge the use of the ASAS-SN data shown in Figure~\ref{fig:lightcurve}.

This work has made use of data from the European Space Agency (ESA) mission
{\it Gaia} (\url{https://www.cosmos.esa.int/gaia}), processed by the {\it Gaia}
Data Processing and Analysis Consortium (DPAC,
\url{https://www.cosmos.esa.int/web/gaia/dpac/consortium}). Funding for the DPAC
has been provided by national institutions, in particular the institutions
participating in the {\it Gaia} Multilateral Agreement.

\facilities{LDT (DeVeny optical spectrograph), Gaia}

\bibliography{masterbib.bib}

\end{document}